**TRIBUNE**

# Research on Research Visibility

*Recerca sobre visibilitat de la recerca*

**Enrique Orduña-Malea**

The iMetrics Lab. Department of Audiovisual Communication, Documentation and History of Art, Universitat Politècnica de València
enorma@upv.es

**Cristina I. Font-Julián**

The iMetrics Lab. Department of Audiovisual Communication, Documentation and History of Art, Universitat Politècnica de València
crifonju@upv.es

## Abstract

This editorial explores the significance of research visibility within the evolving landscape of academic communication, mainly focusing on the role of search engines as online meta-markets shaping the impact of research. With the rapid expansion of scientific output and the increasing reliance on algorithm-driven platforms such as *Google* and *Google Scholar*, the online visibility of scholarly work has become an essential factor in determining its reach and influence. The need for more rigorous research into academic search engine optimization (A-SEO), a field still in its infancy despite its growing relevance, is also discussed, highlighting key challenges in the field, including the lack of robust research methodologies, the skepticism within the academic community regarding the commercialization of science, and the need for standardization in reporting and measurement techniques. This editorial thus invites a multidisciplinary dialogue on the future of research visibility, with significant implications for academic publishing, science communication, research evaluation, and the global scientific ecosystem.

## Resum

Aquest editorial explora la importància de la visibilitat de la recerca en el context canviant de la comunicació acadèmica, centrant-se principalment en el paper dels motors de cerca com a metamercats en línia que modelen l'impacte de la recerca. Amb l'expansió ràpida de la producció científica i la creixent dependència de plataformes impulsades per algoritmes, com ara *Google* i *Google Scholar*, la visibilitat en línia del treball acadèmic s'ha convertit en un factor essencial per determinar el seu abast i influència. També es discuteix la necessitat d'una investigació més rigorosa sobre l'optimització de motors de cerca acadèmics (A-SEO), un camp que encara es troba en les seves primeres etapes malgrat la seva rellevància creixent. Es destaquen els principals reptes del camp, com ara la manca de metodologies d'investigació sòlides, l'escepticisme dins de la comunitat acadèmica respecte a la comercialització de la ciència i la necessitat d'estandardització en les tècniques de mesura i report. Aquest editorial convida així a un diàleg multidisciplinari sobre el futur de la visibilitat de la recerca, amb implicacions significatives per a l'edició acadèmica, la comunicació científica, l'avaluació de la recerca i l'ecosistema científic global.

## Paraules clau

comunicació científica, metaciència, ciència de la ciència, optimització de motors de cerca acadèmics, visibilitat científica, estudis de ciència, cienciometria, informetria

## Funding

Grant PID2022-142569NA-I00, funded by MCIN/AEI/10.13039/501100011033 and by "ERDF A way of making Europe".

## Recommended citation

Orduña-Malea, Enrique, and Font-Julián, Cristina I. (2024). Research on research visibility. *BiD*, 53. https://doi.org/10.1344/bid2024.53.01

# 1. Science as an Object

Science, or the scientific enterprise, is an object of research on its own whose broader understanding is provided by contributions from diverse fields and disciplines under different epistemological assumptions.

In some cases, we find fields exclusively dedicated to studying science, such as meta-research (Ioannidis, 2018), which focuses primarily on how research is conducted, reported, and replicated, or the so-called Science of Science (Fortunato *et al.*, 2018), which is predominantly interested in the dynamics of the production, use, and evaluation of science, which is based fundamentally on the quantitative study of scientific results (objects) by different actors (authors, research institutions, journals, funding and evaluation agencies, etc.). Science studies is also a term increasingly used to cover all the previous aspects, although not always with the same meaning or thematic breadth.

In other cases, we can find specializations of other established disciplines, partially intertwined with the previous fields, such as the philosophy of science (e.g., Rosenberg and McIntyre, 2019), which focuses on the foundational, methodological, and epistemological aspects of sciences; the sociology of science (e.g., Merton, 1973), which aims to understand the social processes and structures that shape scientific knowledge and practice; the history of science (e.g., Kuhn, 2014), which traces the historical context in which scientific discoveries were made and how science has shaped and been shaped by society, politics, and culture; the psychology of science (e.g., Maslow, 1966), which examines the cognitive and psychological processes involved in scientific thinking, creativity, and problem-solving; science policy (e.g., Fealing, 2011), which conducts studies on how governments and organizations design, create and implement policies (including science diplomacy), and how these policies influence scientific research and citizenship; economics of science (e.g., Stephan, 2011), which covers the design and test of diverse funding mechanisms, the division of scientific labor and the economic impact of science on society; and Science communication (e.g., Fischhoff, 2013), which studies how the scientific knowledge is shared, disseminated, promoted, discovered and used, both offline and online.

## 2. Science as a Communicative Object

When studying science (especially the scientific activity) as a communicative object, either theoretically or practically, the studies can be characterized under three different attributes: the object (i.e., what is being measured or studied), the interaction (i.e., with what element does the object interact in the communicative process), and the role (i.e., what role does the element play in the communicative process).

*Object axis.* The focus can be placed on analyzing the communication of scientific knowledge (i.e., the content), which includes the dissemination of theories, hypotheses, methodologies, or findings, or the carrier of said scientific knowledge (i.e., the continent), where research publications (considering all scientific genres) and aggregates (e.g., journals, publishers) stand out. In addition, other studies can focus specifically on the actors responsible for different scientific activity tasks (e.g., authors, reviewers, institutions, funding agencies).

*Interaction axis.* The communication of science can occur between objects belonging to the scientific community (i.e., scholarly communication) or including objects outside this community (i.e., science communication). While the first type of interactions includes bibliometric analyses, studies on publication models (including open access), peer-review, or information sharing in scientific events, the second type consists of the transference of scientific knowledge to business companies, the promotion and dissemination of science among citizens and young people, scientific journalism and public understanding of science. The rise of the web and social networking platforms favored the emergence of "heterogeneous" studies (Costas *et al.*, 2021), where the scientific community uses popular media (personal web pages, social profiles, search engines) to communicate and disseminate research results basically with other researchers. However, citizens and organizations outside the scientific community may participate (and they do) in this communication process. We can find webometrics and altmetrics studies within this third type of interaction.

*Role axis.* The role played by each object in each interaction can also determine the study. When the focus is placed on receivers (consumers of scientific information), the different theories of information search and retrieval and search behavior come into play, for example, the information foraging theory (Pirolli and Card, 1999). In contrast, academic marketing emerges when the focus is on the transmitter (creator). In both cases, the channel used (to disseminate or to find) acts as an information filter. Indeed, the channel can be the object of study (e.g., search engines, social media networking sites, bibliographic databases).

## 3. Science Mediated by Marketplaced Channels

Even though the importance of channels in scholarly and science communication has always existed, their role in communication processes related to science allows us to frame science studies within the scientific economy of attention, under which science is viewed as a "society of producers who pay attention to the information produced by others as an input to the production of their information, which in turn will be usable to others (Franck, 2022). This way, scientific communication' is viewed as a 'market' in which the supply of scientific information meets demand and is exchanged for attention.

This way, scientific communication occurs through channels that act as 'marketplaces', such as conferences, journals, bibliographic databases, search engines, social media platforms, and AI-based conversational search engines. In all these places, "the attention a scientist earns is [capitalized] into the asset called reputation" (Franck, 2022).

The operation of marketplaces affects how science (the content, the author of the content, and the institution of the author) is communicated, perceived, read, trusted, and valued. Arguably, those platforms that concentrate the search for scientific information (i.e., general and academic search engines) can be described as co-producers of academic knowledge (Van Dijck, 2010).

For example, *Google Scholar* surpasses forty million visits each month (Figure 1), considering only the scholar.google.com domain name and the desktop-based traffic. For this reason, the way *Google Scholar* responds to each query affects the research found and eventually read and cited.

Figure 1. Monthly web traffic to *Google Scholar*

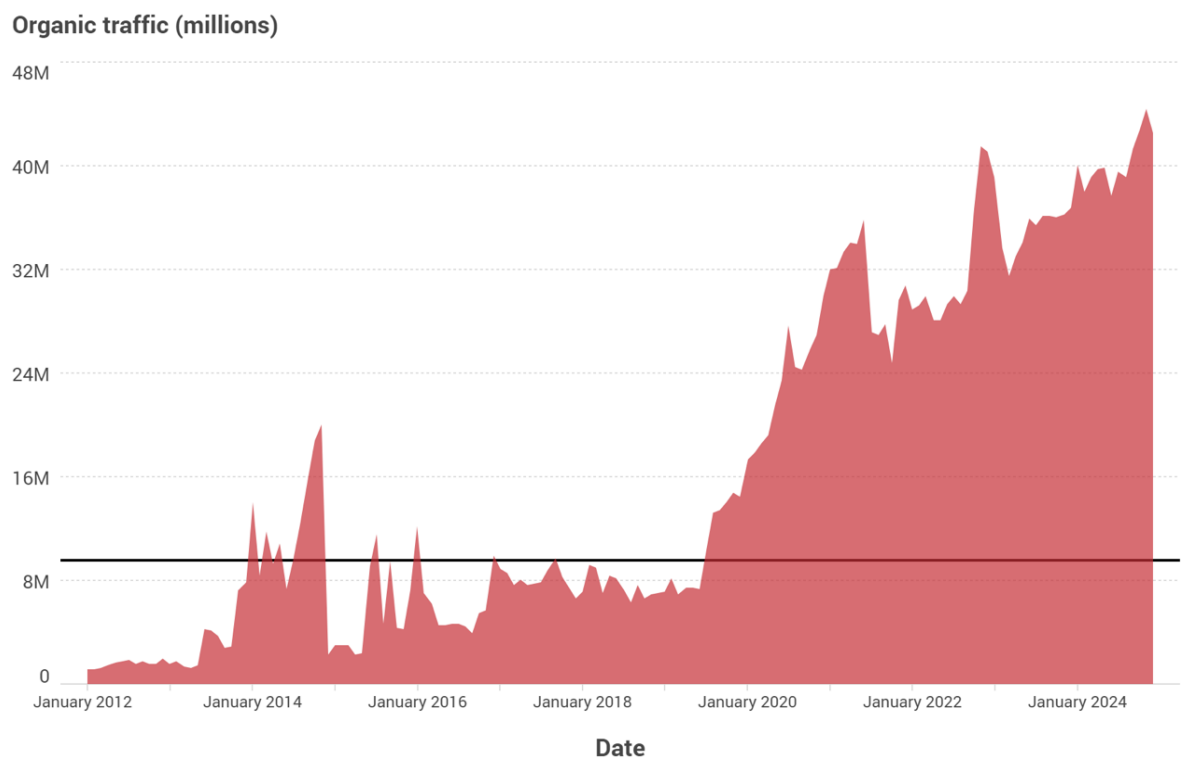

Note: data is limited to desktop-based traffic to scholar.google.com.
Source: *Semrush*. Horizontal line: median value.

The difficulty of getting attention and the reputation gained differ in each marketplace, and this attention depends on being visible to others. The massively used platform-based communication channels are operated by algorithms that respond to economic interests away from the idiosyncrasy of scientific activity (Ma, 2023), making being visible increasingly complex.

Visibility in search engines is limited, fleeting, variable, and subject to the functionalities of the platforms and websites where publications are made accessible, which can change independently of the quality or relevance of a work. Furthermore, visibility depends on how users search for information and browse between the results obtained. These skills rely on discipline, cultural, linguistic, gender, age, among other issues.

The visibility of scientific publications (and their authors and venues of publication) on online marketplaces constitutes a growing field of research, partially covered by altmetrics (Sugimoto *et al.*, 2017), mainly based on social networking sites; webometrics (Thelwall, 2012), based on websites; academic search engine optimization (Beel *et al.*, 2010), based on general and academic search engines; and the upcoming Academic Generative Engine Optimisation (A-GEO) (Aggarwal *et al.*, 2024; Font-Julián *et al.*, 2024; Urbano, 2024), based on large language models (LLMs).

Specifically, the literature on A-SEO has shown, among other aspects, the low indexing rates of publications hosted in institutional repositories (Arlitsch and O'Brien, 2012; Orduña-Malea and Delgado López-Cózar, 2015; Orduña-Malea *et al.*, 2024); the relevance of citations and language in the positioning of bibliographic records in search engine results pages (Martín-Martín *et al.*, 2017; Rovira *et al.*, 2019; 2021), which has, in turn, caused an increase in citations to old publications (Martín-Martín *et al.*, 2016); and the generation of traffic from AI-based search engines to the journals (Urbano, 2024). Other studies have focused on identifying those formal elements that help the findability of publications (Beel *et al.*, 2010) and the growing illicit use of links to and from academic sites to generate visibility on commercial or fraudulent websites, taking advantage of the web reputation of academic sites (Orduña-Malea, 2021). All these findings, together with those already provided by related disciplines, make up a growing body of knowledge about the online visibility of scientific research and glimpse the effects that invisibility can bring.

However, despite the findings, the area of A-SEO is still very scarce, especially compared to webometrics and altmetrics. It is due to the following considerations:

First, despite the growing scientific literature on SEO, unfortunately, many of these works are aimed at offering recipes or recommendations to improve the visibility of the works, their authors, or the journals that publish them. Even though these works are helpful, especially at a professional level, in many cases, they lack robust research methodologies, are excessively descriptive, and resemble what a consultancy report might produce, not a scientific peer-reviewed publication.

Second, and probably as a partial consequence of the previous limitation, it is a field that generates rejection from the research community because it is assumed that it promotes and facilitates a transformation of the natural evolution of publications; that is, it is perceived as a deception and an interference of marketing in the pure and naïve activities of science and researchers. Beel and Gipp (2010), after their pioneering studies on A-SEO, shared negative comments received by evaluators who rejected these practices, as it "seems to encourage scientific paper authors to learn Google scholar's ranking method and write papers accordingly to boost ranking [which is not] acceptable to scientific communities which are supposed to advocate true technical quality/impact instead of ranking." Unfortunately, the scientific literature has evidenced cases of fraud and manipulation of documents and citations in academic databases such as Google Scholar (Delgado López-Cózar *et al.*, 2014).

This rejection does not only occur at the research level, since in the professional environment, but the authors have also encountered during the delivery of courses and lectures the opinions of professionals, mainly in libraries and repositories (both technical and non-technical staff) who expressed their rejection of taking actions aimed at maximizing the visibility of publications since they were not within their competence. Even when the authors understand this argument, it is merely a question of professional roles, not of the relevance of the actions to be developed. The teams in charge of online infrastructures dedicated to making research results accessible must gradually incorporate professionals who implement research visibility strategies into their teams. The consolidation of these professionals would feed back the need for research in the area and would result in an improvement of scientific information systems.

Third, literature lacks standardization in the methodologies, techniques, and indicators used, as well as in reporting the results. These issues, together with the high dynamism of the web and the opacity and variability of search engines' functionalities, limit the comparability and replicability of results, jeopardizing the processes of knowledge accumulation. Otherwise, the authors (including ourselves) use promotion, communication, dissemination, and visibility in different ways, sometimes as synonyms, limiting the performance of systematic reviews that would allow for structuring the knowledge already acquired by the community.

## Final Remarks

This editorial aims to highlight the importance and relevance of the visibility (or invisibility) of search results in meta-markets (search engines, social media platforms, AI-based search engines) in the impact of the works and the reputation of their authors and institutions.

Therefore, it is necessary to strengthen the literature on A-SEO based on understanding the processes of searching, retrieving, and using scientific literature in online channels, for which it is recommended to drive A-SEO studies to the broader research on research visibility, including more conceptual and experimental studies, as well as works aimed to identify, describe, compare and classify data sources, indicators and techniques, and combining quantitative and qualitative methods. The formulation of publication guides would facilitate and structure the reporting of results.

In addition, the launching of specialized journals and a strong presence at consolidated international conferences are commendable, for which expert reviewers are deemed necessary to avoid peer review by non-experts, as is often the case today.

Research visibility should also be part of doctoral training courses for early researchers in any discipline so that they become aware of their importance and role in research work, as well as minimum skills to carry out optimal visibility tasks, since research does not end with publication, but continues with its sharing and dissemination. Likewise, the training of expert research staff in this field is necessary through the formation of specialist re-search groups, the supervision of doctoral theses, or the launching of specialization courses in order to strengthen research work in the area.

All these challenges represent a path for advancing a dynamic and multidisciplinary field of knowledge, with enormous implications in designing funding and evaluation policies for science. For example, both altmetrics and science dissemination activities have begun to have a place among the evaluable merits in the accreditation processes for university teaching staff in Spain through the Spanish National Agency for Assessment and Accreditation (ANECA).

However, this line of research should not be isolated from other areas dedicated to scientific research. Research on research visibility should be embedded and connected with other fields, such as webometrics, altmetrics, bibliometrics, and science communication, shaping a more solid and, above all, more unified line of research.